\documentclass[3p,times,procedia]{elsarticle}
\usepackage{nupha_ecrc}

\volume{00}

\firstpage{1}

\journalname{Nuclear Physics A}

\runauth{}

\jid{nupha}

\jnltitlelogo{Nuclear Physics A}

\usepackage{amssymb}
\usepackage[figuresright]{rotating}

\begin{document}
\begin{frontmatter}



\dochead{}
%% e.g. \dochead{17th International Conference on Dynamical Processes in Excited States of Solids}

\title{Centrality and pseudorapidity dependence of the transverse energy flow in pPb collisions at $\sqrt{s_{\rm NN}} = 5.02$ TeV}


\author{Christopher Bruner and Michael Murray for the CMS Collaboration}
\address{University of Kansas, Lawrence, KS 66045}

\begin{abstract}
The almost hermetic coverage of CMS is used to measure the distribution of transverse energy as a function of pseudo-rapidity  for pPb collisions at  
$\sqrt{s_{NN}} = 5.02$ TeV. For minimum bias collisions  $(1/N)~dE_T/d\eta$ reaches 23 GeV which implies an $E_T$ per participant pair  comparable to that of peripheral PbPb collisions at $\sqrt{s_{NN}} = 2.76$ TeV. 
The centrality dependence of transverse energy production has been studied using centrality measures defined in three different angular regions. 
There is a strong auto-correlation between  $(1/N)~dE_T/d\eta$ and the $\eta$ range used to define centrality %both 
for data and the EPOS-LHC and HIJING event generators.  
The centrality dependence of the data is much stronger  for $\eta$ values on the lead side than the proton side and shows significant differences from that predicted by either event generator. 
\end{abstract}

\end{frontmatter}
%\linenumbers


\vspace{1.0cm}
%\section{Introduction}
The total transverse energy, or $E_T$, produced in a heavy ion or proton nucleus event is a measure of the energy density produced in that event. This energy density is estimated by Bjorken to be
%\begin{equation}
$
\epsilon_{BJ} = \frac{dE_T}{dy}\cdot \frac{1}{\tau_0 A},
$
%\label{Eqn:EnergyDensity}
%\end{equation}
%For high energy collisions  $\frac{dE_T}{dy} \approx  \frac{dE_T}{d\eta}$.
where A is the transverse area of the nuclear overlap zone and $\tau_0$ represents a time. 
For $\sqrt{s_{NN}} = 2.76 $ TeV PbPb collisions energy densities up to 14 GeV/fm$^3$ are observed \cite{Chatrchyan:2012mb}, assuming $\tau_0 = 1$ fm/c. This value is much higher than the threshold for the production of a quark gluon plasma. It has been suggested that collective phenomena such as azimuthal flow have been observed in pPb collisions. It is interesting to see if large energy densities are also observed in pPb collisions.
 When using a particular $\eta$ region to define centrality there is obviously an auto correlation with the transverse energy $dE_T/d\eta$ measured in that $\eta$ region. 
%For symmetric heavy ion collisions these auto-correlations are not large 
%\cite{Chatrchyan:2012mb}. 
%This is not expected to be the case for the small system created in pPb collisions.  
% For PbPb collisions their is only a weak 
%$\eta$ dependence of the centrality dependence \cite{Chatrchyan:2012mb}. 
The very large angular coverage of CMS allows for the study of these effects 
over a very wide $\eta$ range and  compare them to predictions from event generators. Full details of the analysis are in Ref \cite{CMS-PAS-HIN-14-014}.

%\section{Trigger and Event Selection}
%\label{Sec:EventSelection}

%The data for this analysis was recorded during the LHC 2013 pPb, and Pbp runs. %During this run 31 nb$^{-1}$ of data are taken by CMS, of which
This analysis is based upon  
1.14 nb$^{-1}$ of pPb data from the  LHC 2013 run. A sample of Pbp events was used to check the symmetry of the calorimeters.
For this analysis  the hardware-based Level 1 zero bias trigger required only that two beams be present in CMS. The High Level Trigger then reconstructs these events  and requires at least one track with $|\eta| < 2.5$  and $p_T > 0.4 $GeV/c. This guarantees that a collision occurred.
%The first requirement of the offline analysis is to select nuclear pPb interactions. 
Offline cuts are made in order to ensure that the collision is a nuclear pPb collision rather than a beam gas or electromagnetic interaction.  
The noise is estimated from a sample of empty events collected 
using the coincidence of a random trigger with the requirement that no beams be present. CMS defines positive $\eta$ as the proton going direction. %This implies that negative $\eta$ is the lead going direction. 
 A  detailed description of  CMS 
 can be found in Ref.~\cite{Chatrchyan:2008zzk}. 

\section{Analysis}
This analysis is based upon objects produced by  the CMS Particle-Flow algorithm  %CiteParticleFlow[36,37].  
which uses information from the tracker, calorimeters and muon chambers to identify and reconstruct individual particles from collisions.
 The transverse energy density  is calculated using 
 the following equation
\begin{equation}
\frac{1}{N} \frac{dE_T}{d\eta}(\eta) = \frac{1}{N}\frac{1}{\Delta \eta} 
 \sum_{j} E_T^{j} ({\rm if}~E_T^{j} > noise)  \cdot C(\eta),
\label{Eqn:EtCal}
\end{equation}
where $E_T^j$ is the $E_T$ of a particular particle flow object, N is the number of good events and $C(\eta)$ %is a correction factor that 
accounts for the acceptance of CMS. The correction is deduced from Monte Carlo simulations.
 and  is defined as
\begin{equation}
C(\eta) = \frac{\sum_k E_T^k ({\rm generated})}{\sum_j E_T^j ({\rm reconstructed}) ({\rm if}~E_T^{j} > noise)}. 
\label{Eqn:CorrectionFactor}
\end{equation}
Correction factors were generated using the  EPOS-LHC and HIJING generators\cite{Pierog:2013ria,Wang:1991hta}. 
The main sources of systematic error on $(1/N)~dE_T/d\eta$ are:
\begin{enumerate}
\item Differences in spectra and particle composition between data and the Monte Carlos;% used to generate correction factors; 
\item Different ways of handling the noise;
\item Any asymmetries between the positive and negative sides of CMS;
%\item Residual pileup that survives event cuts. 
\item Uncertainties in the calorimeter energy scale;
\end{enumerate}
For the analysis of the reconstructed Monte Carlo events, the event selectioncuts are the same as for the data.  
No noise cuts are applied to the Monte Carlo events. The corrections are calculated by using EPOS-LHC events with that have been weighted to have the same $E_T$ per charged hadron as the data.  Systematics due to the slight differences in particle composition between EPOS-LHC and data were estimated using correction factors derived from HIJING events. 
The systematic errors as a function of centrality are listed in Table \ref{Tab:SysEtTracker}. 
\begin{table}[h]
\begin{center}
\begin{tabular}{|c|c|c|c|c|c|} \hline
Centrality                &    MC & Noise & pPb -Pbp & E scale & Total \\  \hline 
%(HF+ and HF-) &   (\%)   &   (\%) &  (\%) &  (\%) & (\%) \\ \hline
Min. bias & 2.7 & 1.7 & 2.4 & 1.0 & 4.2 \\ \hline
0  --10\%        &  2.9 & 0.9  & 1.8 & 1.0 & 3.7 \\ \hline
10 -- 20\%      &  2.8 & 1.1  & 2.0 & 1.0 & 3.8 \\ \hline
20 -- 30\%      &  2.7 & 1.3  & 2.1 & 1.0 & 3.8 \\ \hline
30 -- 40\%      &  2.6 & 1.5  & 2.1 & 1.0 & 3.8 \\ \hline
40 -- 50\%      &  2.5 & 1.8  & 2.7 & 1.0  & 4.2 \\ \hline
50 -- 60\%      &  2.4 & 2.3  & 2.9 & 1.0  & 4.5 \\ \hline
60 -- 70\%      &  2.3 & 3.1  & 3.2 & 1.0  & 5.1 \\ \hline
70 -- 80\%      &  2.2 & 4.2  & 5.0 & 1.0  & 7.0 \\ \hline
\end{tabular}
\end{center}
\caption{Systematic errors (\%) on $(1/N)~dE_T/d\eta$.  There is a significant correlation between the errors at different centralities. } 
\label{Tab:SysEtTracker}
\end{table}

\section{Results} 
\begin{figure}[h]
\centering
\includegraphics[width=0.4\columnwidth]{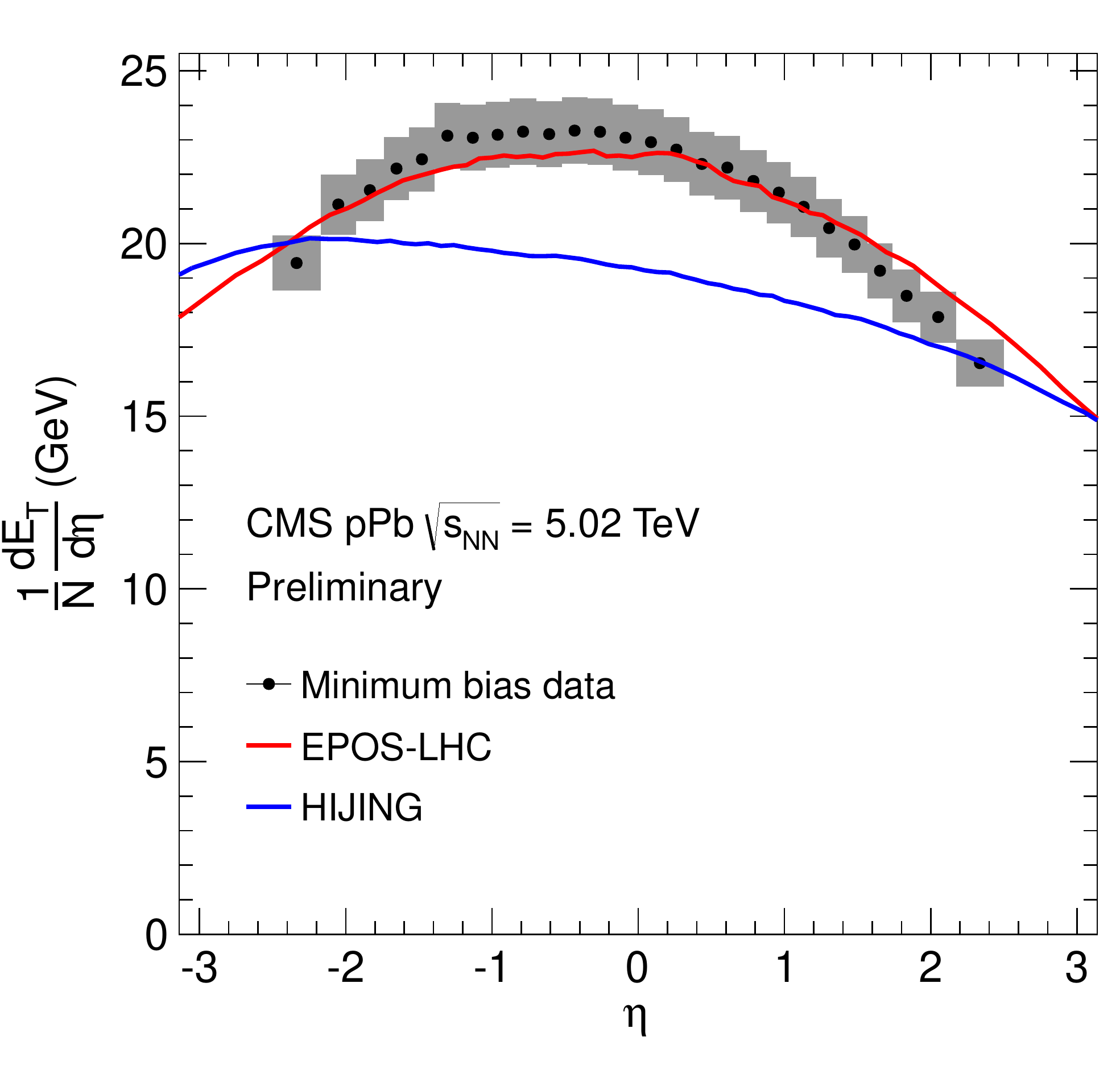}
\includegraphics[width=0.4\columnwidth]{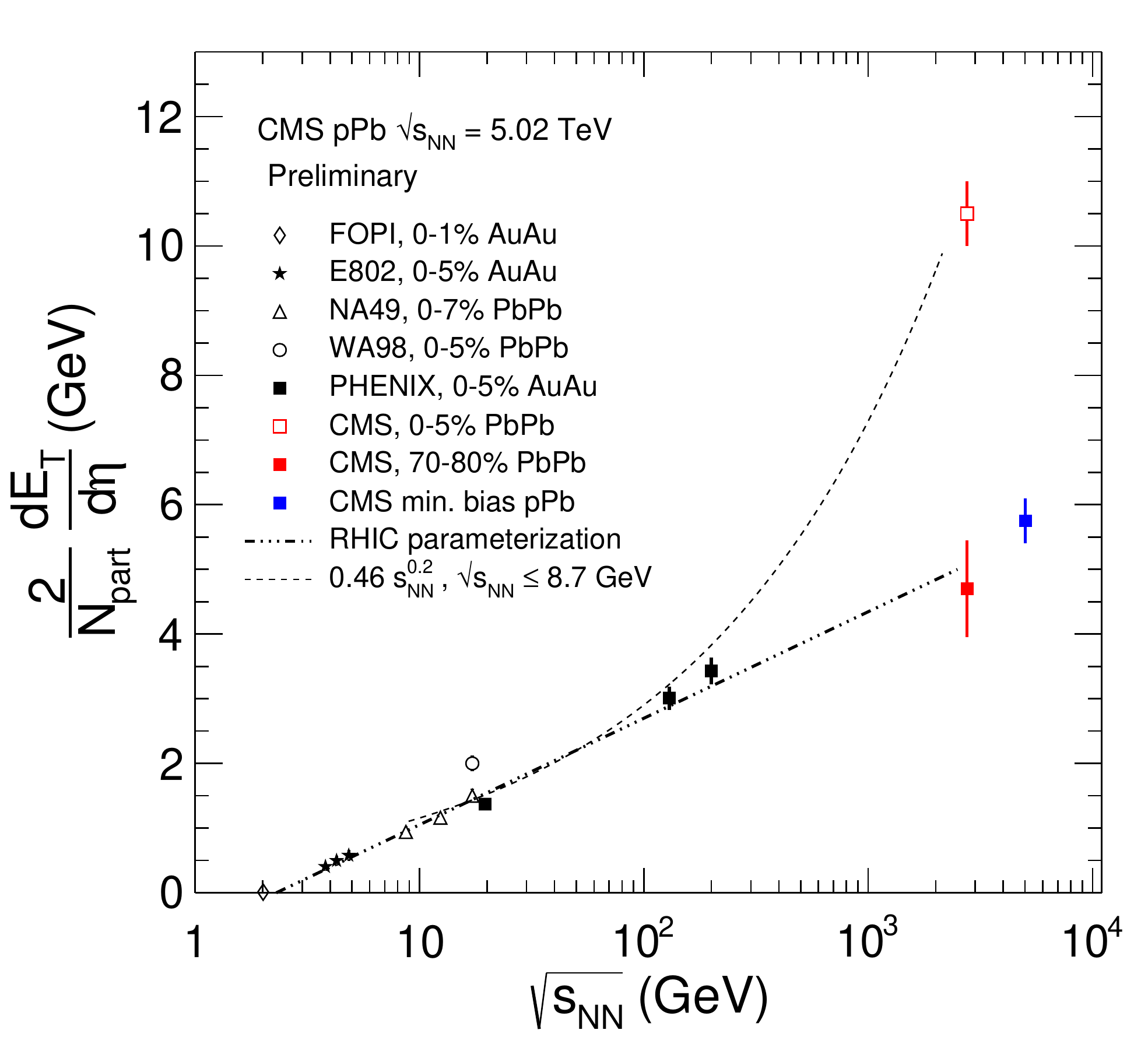}
\caption{(Right) $(1/N)~dE_T/d\eta$ versus $\eta$ from minimum bias $\sqrt{S_{\rm NN}} = 5.02 $TeV pPb collisions. The statistical errors are smaller than the size of the points. The systematic errors are shown by gray bands and are largely correlated point to point.   Predictions from EPOS-LHC (red) and HIJING (blue) are also shown. 
(Left) $dE_T/d\eta \cdot 1/(0.5 \cdot N_{\rm part})$ versus $\sqrt{S_{\rm NN}}$ for AuAu, PbPb and pPb collisions 
%  \cite{Chatrchyan:2012mb,Adler:2004zn,Abbott:2001gd,Aggarwal:2000bc,Bachler:1999hu,Afanasiev:2002fk,Afanasiev:2002mx,Reisdorf:1900zza,Pelte:1997rg,Hong:2001tm}. 
The RHIC parametrization (dashed-dotted) is extended to higher energies to guide the eye.
\label{Corrected_minbias_data}}
\end{figure}

Figure \ref{Corrected_minbias_data}  shows the corrected 5.02 TeV pPb minimum bias 
$(1/N)~dE_T/d\eta$ versus $\eta$ and centrality  for data, EPOS-LHC and HIJING.  
At $\eta=0,$  $(1/N)~dE_T/d\eta$ is 23 GeV. This is 1/40 times the value observed for central PbPb collisions. However we expect that the cross sectional area of a pPb collision should is of the order 50 times smaller than for PbPb. This implies that the energy density in pPb collisions is comparable to that achieved in PbPb. 
EPOS-LHC is consistent with the data over almost the whole region while HIJING is both below the data and is peaked more on the Pb side than the data.   
The right hand side of Fig.~\ref{Corrected_minbias_data} 
compares the pPb 5.02 TeV results to lower energy heavy ion results, \cite{Chatrchyan:2012mb}. To account for the different system sizes the  $(1/N)~dE_T/d\eta$ values are normalized to the number of participants in the collisions.  
%For min. bias pPb there are approximately 8 participating  nucleons compared to 16 for peripheral PbPb. For central PbPb collisions the $N_{\rm part}$ can reach almost 400.  
The  pPb value of  
$\approx 5.8 $GeV per participant pair   
is consistent with the peripheral PbPb result at  $\sqrt{S_{\rm NN}} = 2.76$ TeV \cite{Chatrchyan:2012mb}.

%\section{Centrality}
For this paper three different measures of centrality are used;
\begin{itemize}
  \item HF-Single: the $E_T$ deposited in lead going side of HF with
  $-5.0<\eta <-4.0$,
  \item HF-Double the $E_T$ deposited in both sides of HF with $4.0<|\eta| < 5.0$, and
  \item  $N_{\rm Track}$: the number of offline tracks with $p_T > 400$ MeV and $|\eta| < 2.4$
\end{itemize}

\begin{figure}[h]
\centering
\includegraphics[width=0.48\columnwidth]{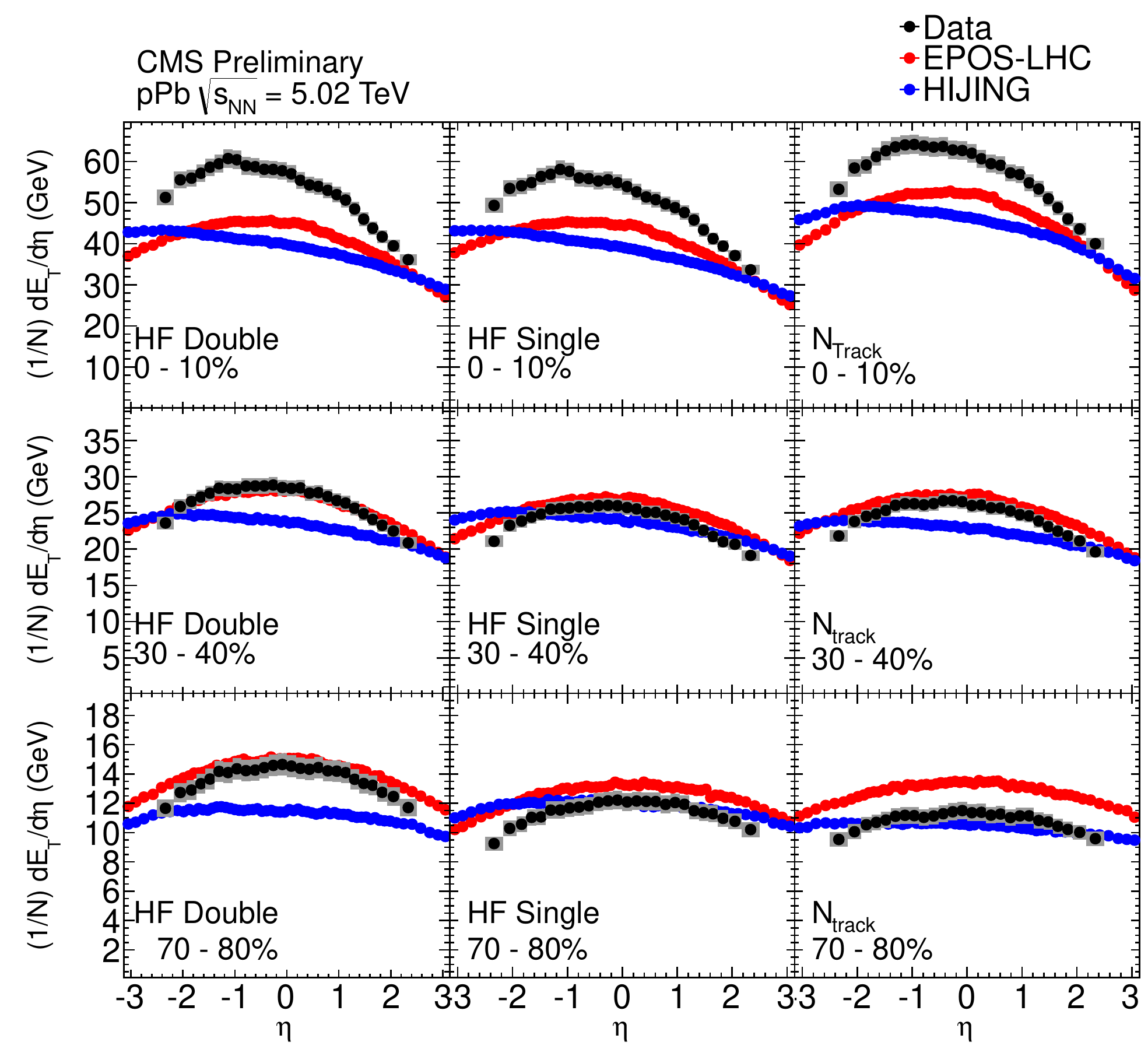}
\includegraphics[width=0.48\columnwidth]{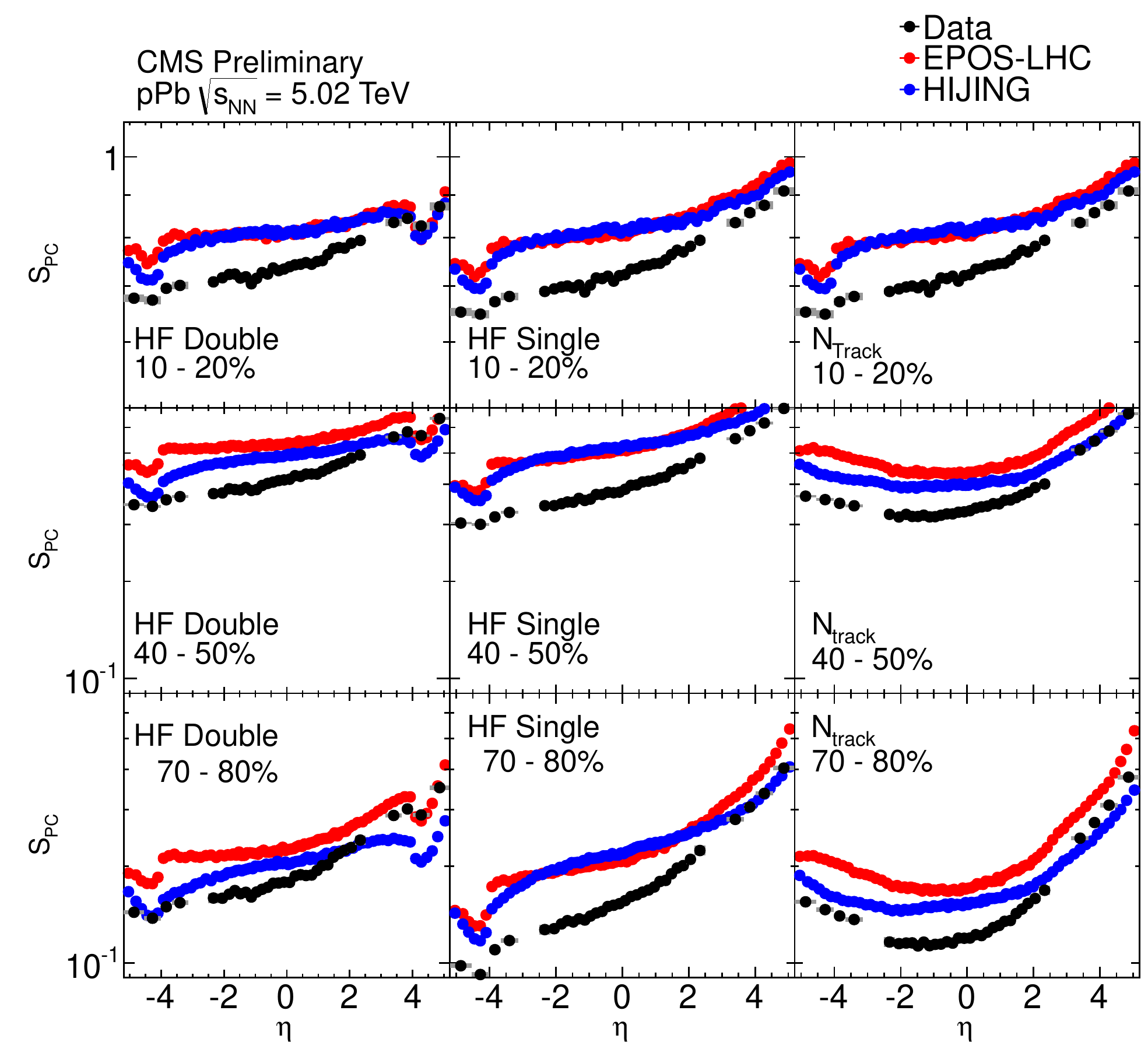}
\caption{$(1/N)~dE_T/d\eta$ (Right) and S$_{\rm PC}$ (Left) versus $\eta$ and centrality  from 5.02 TeV pPb collisions for
HF-double (left), HF-single (center), and $N_{\rm Track}$ (right) centrality definitions for data EPOS-LHC  and HIJING, for central collisions (top), 30 - 40\% (middle), and 70-80\% (bottom). 
  The gray bands show the systematic errors on the data. These are largely correlated point to point.
  \label{Eflow_Centrality_Comparison}}
\end{figure}

Figure \ref{Eflow_Centrality_Comparison} (Left) 
shows $(1/N)~dE_T/d\eta$ versus $\eta$ and centrality for pPb collisions selected with three centrality definitions. As the centrality increases $(1/N)~dE_T/d\eta$  at $\eta = 0$ increases rapidly and the peak moves backwards in $\eta$. Near $\eta=0$ the increase with centrality is stronger for the $N_{\rm Track}$ centrality definition.
%at $\sqrt{S_{NN}} = 5.02$ TeV for the HF-double, HF-single, and $N_{\rm Track}$ centrality definitions, for data, EPOS-LHC and HIJING. 
As the centrality increases $(1/N)~dE_T/d\eta$  at $\eta = 0$ increases faster for the  $N_{\rm Track}$ centrality definition than for the HF-single or HF-double definitions. 
%a factor of 20 or so for the  HF-double and HF-single centrality definitions and by a factor of 30 for the $N_{\rm Track}$ centrality definition. 
%The greater range of the data for the  $N_{\rm Track}$ centrality definition reflects the auto-correlation from the centrality definition.  
For the HF-double $(1/N)~dE_T/d\eta$  reaches 60 GeV at $\eta=0$ implying a very high energy density in such collisions.  
Both EPOS-LHC and HIJING show a large increase of $(1/N)~dE_T/d\eta$ and a shift toward in $\eta$ towards the Pb side as the centrality increases. EPOS-LHC does  better at reproducing the shape of the data than does HIJING.  As in the data $(1/N)~dE_T/d\eta$ near $\eta=0$ is larger for more central events selected by $N_{\rm Track}$ than for the cases where the centrality is deduced from HF.  

%In order to understand the fluctuations and autocorrelations that can arise from to a choice of centrality, a measurement needs to be made over as large an $\eta$ range as possible for several different centrality definitions. 
In order to focus on the $\eta$ dependence of the centrality dependence
% of the data as a function of $\eta$, the ratio of peripheral to central $\frac{1}{N} \frac{dE_T}{d\eta}(\eta)$  
the ratio S$_{\rm PC}$ is defined as
\begin{equation}
{\rm S}_{\rm PC} (\eta)= \frac{\frac{dE_T}{d\eta}(peripheral,\eta)}{\frac{dE_T}{d\eta}(central,\eta)}  = \frac{\sum_{i} E_T^{i} (peripheral)}
{\sum_{i} E_T^{i} (central)}  \cdot \frac{C(peripheral,\eta)}
{C(central,\eta)}.
\label{Eqn:SpcDef}
\end{equation}
Since S$_{\rm PC}$  depends upon the ratio of data samples and correction factors,  correlated errors tend to cancel. S$_{\rm PC}$  is by construction equal to unity for central events and decreases for more peripheral events.

Figure~\ref{Eflow_Centrality_Comparison} (Right) 
shows  S$_{\rm PC}$  as a function of $\eta$ for 3 centrality ranges and for
all three centrality definitions for data EPOS-LHC and HIJING. S$_{\rm PC}$ tends to rise with $\eta$, i.e. as one moves from the lead to the proton rapidity since the centrality dependence of  $dE_T/d\eta$ is stronger on the lead side. 
%For all centrality definitions and at all values of $\eta$,  S$_{\rm PC}$  decreases going from central to peripheral collisions. 
The effect of the autocorrelations is a dip in the $\eta$ region 
where the centrality is defined. The autocorrelation in $dE_T/d\eta$  causes a dip in the peripheral distribution but a bump in
the central distribution and so the ratio of peripheral/central shows a dip.
This dip in S$_{\rm PC}$ is  larger for more peripheral
events because the effect of the centrality selection is more pronounced.  
For HF-double and HF-Single S$_{\rm PC}$ tends to fall with $\eta$ but this is not the case when $N_{\rm Track}$ is used to define centrality. For this definition of centrality S$_{\rm PC}$ is smallest near $\eta=1$.

 For both EPOS and HIJING the general shapes of the curves are the same for the data and the models but S$_{\rm PC}$ is smaller for the most peripheral data than in the models. For the HF-double definition of centrality both models 
 %EPOS-LHC and HIJING 
 show a more pronounced dip on the proton side than is seen in the data. This implies that the models predict a larger change in $E_T$ with centrality for the proton region than is seen in the data.

\section{Summary}
Proton lead  collisions at $\sqrt{s_{\rm NN}} = 5.02$ TeV produce a large transverse energy over a very wide pseudo-rapidity range. 
The transverse energy per participant pair in minimum bias pPb collisions at $\sqrt{s_{\rm NN}} = 5.02$ TeV is comparable to that of peripheral 
PbPb collisions at  $\sqrt{s_{\rm NN}} = 2.76$ TeV. 
%This implies that the energy densities in such collisions are comparable to those of PbPb.  
 As the centrality of the collision increases the total $E_T$ increases dramatically and the  mean $\eta$ of the $E_T$ moves towards the lead side. 
 The EPOS-LHC event generator gives a good description of the minimum bias $dE_T/d\eta$ distribution and  peaks at an $\eta$ value close to that of the data for all centralities. However EPOS-LHC under predicts the magnitude $dE_T/d\eta$ for central collisions. HIJING consistently under predicts the magnitude of  $dE_T/d\eta$ and has a peak that significantly to the left of the data for central collisions.   
The $\eta$ region used to define centrality has a strong influence on the nature of the events selected.  
There is a strong auto-correlation between  $(1/N)~dE_T/d\eta$ and the $\eta$ range used to define centrality both for data and the EPOS-LHC and HIJING event generators.  
The centrality dependence of the data is much stronger  for $\eta$ values on the lead side than the proton side and shows significant differences from that predicted by either of the event generators.



%% References


\bibliographystyle{elsarticle-num}
\bibliography{MurrayQM15Refs}
\end{document}